\documentclass[aps,prc ,twocolumn, nofootinbib,superscriptaddress]{revtex4-1}

\usepackage{simpler-wick}
\usepackage{graphicx}  
\usepackage{amsmath}
\usepackage{amsfonts}
\usepackage{amsbsy}
\usepackage{bm}  %
\usepackage{color}
\usepackage{tikz}
\usepackage{xcolor}
\usepackage{hyperref}
\usepackage{float}
\usepackage{xspace}

\usepackage{ulem}

\hypersetup{
    colorlinks,
    linkcolor={red!50!black},
    citecolor={blue!50!black},
    urlcolor={blue!80!black}
}

\newcommand{\bea}{\begin{eqnarray}}
\newcommand{\eea}{\end{eqnarray}}
\newcommand{\beq}{\begin{equation}}
\newcommand{\eeq}{\end{equation}}
\newcommand{\be}{\begin{equation}}
\newcommand{\ee}{\end{equation}}
\newcommand{\bqa}{\begin{eqnarray}}
\newcommand{\eqa}{\end{eqnarray}}

\def\mqo2{{\!\!\!}}

\newcommand*{\Mlo}{\ensuremath{\Lambda_l}\xspace}
\newcommand*{\Mhi}{\ensuremath{\Lambda_B}\xspace}

\newcommand*{\Mpi}{\ensuremath{M_\pi}\xspace}
\newcommand*{\MN}{\ensuremath{m_N}\xspace}

\newcommand{\prob}{p}

\begin{document}
% * Title
%\title{Determining the breakdown scale of chiral perturbation theory in the one-nucleon sector}

%\title{Inferring the breakdown scales of $g_A$ and \MN in chiral perturbation theory}
\title{Inferring the breakdown scales of the chiral expansions for $g_A$ and \MN}
\author{Andreas Ekström}\email{andreas.ekstrom@chalmers.se}
\affiliation{Department of Physics, Chalmers University of Technology, SE-412 96, Göteborg, Sweden}

\author{Daniel R. Phillips}\email{phillid1@ohio.edu}
\affiliation{Institute of Nuclear and Particle Physics and Department of Physics and Astronomy,\\ Ohio University, Athens, Ohio 45701, USA}

\author{Lucas Platter}\email{lplatter@utk.edu}
\affiliation{Department of Physics and Astronomy, University of Tennessee, Knoxville, Tennessee 37996, USA}
\affiliation{Physics Division, Oak Ridge National Laboratory, Oak Ridge, Tennessee 37831, USA}

\author{Matthias R. Schindler}\email{mschindl@mailbox.sc.edu}
\affiliation{Department of Physics and Astronomy, University of South Carolina, Columbia, South Carolina 29208, USA}
\date{\today}

\begin{abstract}
We apply Bayesian inference to the order-by-order chiral perturbation theory ($\chi$PT) expansions for the axial-vector coupling constant $g_A$ and the nucleon mass \MN, and thereby infer the scales at which $\chi$PT breaks down for these two observables. Using a pointwise Bayesian analysis, we find that the inferred breakdown scales are notably different for the two observables. For the chiral expansion of $g_A$, we obtain $251^{+20}_{-50}$~MeV and $211^{+20}_{-30}$~MeV using two distinct sets of low-energy constants, while for the chiral expansion of \MN{} we infer a significantly larger breakdown scale of $491^{+60}_{-90}$~MeV.
\end{abstract}

\smallskip
\maketitle
%\tableofcontents
\section{Introduction}
\label{sec:intro}

%{\bf [mrs: we need to decide on a notation for the pion mass, also $\Lambda_B$ vs $M_{hi}$], not $\Lambda_{\chi}$}

Chiral perturbation theory ($\chi$PT) \cite{Weinberg:1978kz,Gasser:1983yg,Gasser:1984gg} is among the most successful effective field theories in nuclear and particle physics. $\chi$PT is based on the fact that pions are Goldstone bosons, associated with the spontaneously broken $SU(2)_L\times SU(2)_R$ chiral symmetry of quantum chromodynamics (QCD) that emerges in the limit of zero up and down quark masses---the chiral limit. %(referred to as the chiral limit). 
The pions' finite mass arises from explicit chiral symmetry breaking due to nonvanishing quark masses $m_q$. 
%\sout{All QCD observables can thus be}
In $\chi$PT observables are expanded in powers of a ratio $Q$ of light to heavy scales, $Q=\Mlo/\Mhi$, 
%[mrs: am happy to use the notation from earlier papers with $M_{hi}$ etc if you prefer]
where the light scale \Mlo is identified with the pion mass $M$ or a momentum $p$ of similar size, and the heavy scale \Mhi denotes the breakdown scale of $\chi$PT. The pion mass itself has such an expansion, where $M$ denotes the leading-order expression of the pion mass with $M^2\propto m_q$. In the mesonic sector 
%\sout{it is typically argued that}
\Mhi is typically assumed to be either $\sim 4 \pi F_\pi$ (where $F_\pi = 92.7\text{ MeV}$ is the pion decay constant)~\cite{Manohar:1983md} or the mass $m_\rho=0.770$ GeV of the $\rho$ meson, 
%\sout{with the latter estimate based on the $\rho$-meson being}
the lightest QCD excitation not explicitly included in mesonic $\chi$PT.

This framework has also been extended to include nucleons. The non-vanishing of the nucleon mass in the chiral limit introduces additional complexities compared to the meson sector \cite{Gasser:1987rb}. Different approaches to these issues have been proposed \cite{Jenkins:1990jv,Bernard:1992qa,Becher:1999he,Gegelia:1999gf,Fuchs:2003qc}; for reviews see, e.g., Refs.~\cite{Bernard:1992qa,Bernard:2007zu,Scherer:2012xha}. Here we focus on the formulation of nucleons as heavy baryons \cite{Jenkins:1990jv,Bernard:1991rq}. 
%\sout{In this approach the leading-order result has the nucleon}
At leading order in this approach, the nucleon is treated as a static field, which acts as a source of Goldstone bosons. Nucleon-pion interactions  are then described by an effective Lagrangian which contains increasing powers of momenta and 
%\sout{quark-masses (with $m_q$ counted as $\sim p^2$)}
the pion mass, leading to expansions of nucleonic observables in powers of $M$ and $p$. Prominent examples include the nucleon mass, the nucleon axial-vector coupling, nucleon form factors, and nucleon polarizabilities~\cite{Bernard:1995dp}. 

Such expansions are of particular importance in lattice QCD, where simulations are frequently performed at pion masses larger than the physical value. Extrapolating these results to the physical point relies on the convergence of the $\chi$PT series. And, even in cases where lattice simulations are performed at the physical pion mass, it may still be advantageous to combine results from simulations carried out at different quark masses.

However, in the one-nucleon sector, the situation is complicated by the appearance of additional scales. In particular, the $\Delta(1232)$ resonance is the lowest-lying (non-pionic) QCD excitation in that sector, and 
%\sout{there has been a long debate about whether it should be explicitly included in the low-energy EFT of single-nucleon properties or not} 
while baryon $\chi$PT without explicit $\Delta$ degrees of freedom is commonly used, arguments for the inclusion of $\Delta$ excitations as active degrees of freedom have been presented since the advent of baryon $\chi$PT~\cite{Jenkins:1990jv,Jenkins:1991ne,Hemmert:1996xg,Hemmert:1997ye}. 

A prominent example for the need of including the $\Delta$ is the chiral expansion of the axial-vector coupling constant $g_A$: the first nontrivial correction enters at relative order $M^2$, the next correction proportional to $M^3$ is numerically very large. This slow convergence has long been attributed to the presence of $\Delta$ excitations inside the loop corrections to $g_A$ ~\cite{Jenkins:1991es,Luty:1993gi,Flores-Mendieta:2000ljq,Hemmert:2003cb,CalleCordon:2012xz,Hall:2025ytt}. And indeed, such inclusion is mandatory in the large-$N_c$ limit of QCD, in order to preserve unitarity~\cite{Dashen:1993jt,Flores-Mendieta:2000ljq}. This introduces the $\Delta$-nucleon mass difference as an additional  scale of approximately 290~MeV---much lower than either $4 \pi F_\pi$ or $m_\rho$---into the expansion.

These considerations naturally lead us to ask: What information about the single-nucleon $\chi$PT breakdown scale or scales can be extracted from order-by-order expansions of observables in that theory? 
Earlier work on pion-nucleon scattering, \MN, and $g_A$ \cite{Djukanovic:2006xc,McGovern:2006fm,Schindler:2006ha,Bernard:2006te} addressed this question by comparing individual higher-order contributions or the resummation of particular subsets of higher-order contributions to observables with lower-order terms. In general, it was found that baryon $\chi$PT expansions cease to be reliable around pion masses of $300\text{ MeV}$ to $500\text{ MeV}$, depending on the observable.

Recently, Bayesian statistical techniques have been developed to analyze the convergence pattern of effective field theories~\cite{Furnstahl:2015rha,Melendez:2019izc}, enabling quantitative estimates of their respective breakdown scales. In Refs.~\cite{Melendez:2017phj,Wesolowski:2021cni,Drischler:2020hwi,Millican:2024yuz,Millican:2025sdp,McClung:2025rtc} these methods were applied to demonstrate that the breakdown scale of chiral EFT, applied to two-nucleon scattering, nuclei with mass number $A=3$ and $4$, and neutron matter, has a most likely value of $\Mhi=600$ MeV. In Refs.~\cite{Ekstrom:2024dqr,Ekstrom:2025ncs} these methods were applied to pionless EFT for two- and three-nucleon systems and a most likely breakdown scale of approximately the pion mass was inferred (but cf.~Ref.~\cite{Bub:2024gyz} for a different perspective). 

In this work, we apply these statistical techniques to single-nucleon $\chi$PT to infer the underlying breakdown scales of the $\chi$PT expansions of the nucleon mass \MN and the axial-vector coupling constant $g_A$. 
We achieve this as follows. First, we introduce the chiral expansions for \MN and $g_A$, see Sec.~\ref{sec:chiralexpansions}. The Bayesian methodology employed to analyze their respective convergence patterns and the inference of the associated breakdown scales are described in Sec.~\ref{sec:results}. We conclude with a discussion of our results and their implications in Sec.~\ref{sec:summary}.

%Determine the BChPT breakdown scale $\Lambda_\chi$ from the chiral expansions of quantities like the nucleon mass \MN and the axial coupling $g_A$. 
%These expansions are known to at least 4th order, so we should get some handle on $\Lambda_\chi$. 
%From \href{https://www.int.washington.edu/sites/default/files/schedule_session_files/M_Hoferichter%20slides.pdf}{Martin Hoferichter's INT talk} it seems clear that using $\Lchi \approx 4 \pi F_\pi$ is a choice that results in very large coefficients for the higher-order contributions.
%Since they are evaluated at fixed momentum, this should also avoid some complications. 

 %We will use Bayesian inference to quantify the probability (density) $p(\Lchi|y,I)$, where $y$ (generically) denotes order-by-order data for $g_A$ \emph{and} \MN, and $I$ is any other information we utilize~\cite{Melendez:2019izc}. It might be valuable to study two quantities and explore the consistency of $p(\Lchi|y,I)$, in addition to consistency across BChPT orders.

%\subsection{Possible scope/extensions}
%Once the formalism is set up, there are several points we could consider:
%\begin{itemize}
%    \item Heavy baryon vs covariant
%    \item no $\Delta$ vs BChPT with $\Delta$s
%    \item \ldots?
%\end{itemize}

\section{Chiral expansions of \MN and $g_A$}
\label{sec:chiralexpansions}

We use established expressions for the $\chi$PT expansions of \MN and $g_A$ as outlined below. The coefficients of these expansions generally depend on the pion-decay constant $F$, the nucleon mass $\mathring{m}$, and the nucleon axial-vector coupling $\mathring{g}$ in their respective chiral limits, and a number of other low-energy constants (LECs): $l_i$ from the fourth-order pionic Lagrangian and $c_i, d_i, e_i$ from the pion-nucleon Lagrangian at second, third, and fourth order, respectively. Renormalized LECs are denoted by a superscript $r$ and depend on the renormalization scale $\mu$. 
The renormalized fourth-order mesonic LECs $l_i^r(\mu)$ are related to the scale-independent LECs $\bar{l}_i$ by 
\begin{equation}
    l_i^r(\mu) = \frac{\beta_{l_i}}{32\pi^2}\left[\bar{l}_i+ 2
    \ln\left(\frac{\Mpi^{(\rm phys) }}{\mu}\right)\right]~,
\end{equation}
with $\beta_{l_3}=-\frac{1}{2}$ and $\beta_{l_4}=2$ contributing in the following analyses.
Analogously, the pion-nucleon LECs $d_i^r(\mu)$ are connected to the scale-independent LECs $\bar{d}_i$ by the expressions \cite{Gasser_2002}
\begin{equation}
    \label{eq:dr-dbar}
    d_i^r(\mu) = \bar{d}_i+ \frac{\beta_{d_i}}{(4\pi F)^2} 
    \ln\left(\frac{\Mpi^{(\rm phys) }}{\mu}\right)~,
\end{equation}
with the $\beta_{d_i}$ given in Ref.~\cite{Siemens:2016hdi}.
For the physical pion mass we take $\Mpi^{(\rm phys) } = 0.139$~GeV. For the pion-nucleon LECs we use the values listed in Table.~\ref{tbl:sets}. They are obtained from a $\chi$PT analysis of the processes $\pi N\to \pi N$ and $\pi N \to \pi\pi N$ in Ref.~\cite{Siemens:2017opr}. The values referred to as ``Set 1" and ``Set 2" in Ref.~\cite{Bernard:2025gto} are the ones we employ in the chiral expansion of $g_A$. The LEC values used in the expansion for \MN are referred to as ``Set $m_N$" and are taken from Ref.~\cite{Hoferichter:2015hva}, with the value of the fourth-order pion-nucleon LEC $\bar{e}_1$ provided in Ref.~\cite{Meissner:2022cbi}.

\begin{table}[t]
\begin{tabular}{ |p{2.5cm}||p{1.5cm}|p{1.5cm}| p{1.5cm} |}
 \hline
 \multicolumn{4}{|c|}{Low-energy constants} \\
 \hline
 LEC & Set 1  & Set 2 & Set \MN\\
 \hline
 $c_1$ [GeV${}^{-1}$]   & -    & - & -1.11\\
$c_2$ [GeV${}^{-1}$]   & 3.51    & 4.89 & 3.13\\
 $c_3$ [GeV${}^{-1}$] &   -6.63& -7.26& -5.61\\
 $c_4$ [GeV${}^{-1}$]  & 4.01 & 4.74 & -\\
 $(\bar{d}_1 +\bar{d}_2)$~[GeV$^{-2}$]    & 4.73 & 3.39 & -\\
 $\bar{d}_{10}$~[GeV$^{-2}$]&   -0.8  & 10.9 & -\\
 $\bar{d}_{11}$~[GeV$^{-2}$]&  -15.6  & -30.9 & -\\
 $\bar{d}_{12}$~[GeV$^{-2}$]& 5.9 & -10.9 & -\\
 $\bar{d}_{13}$~[GeV$^{-2}$]& 13.6 & 27.7 & -\\
 $\bar{d}_{14}$~[GeV$^{-2}$]& -7.43 & -7.36 & -\\ 
 $\bar{d}_{16}$~[GeV$^{-2}$]& 0.4 & -3.0 & -\\
 $\bar{d}_{18}$~[GeV$^{-2}$]& -0.8 & -0.8 & -\\
 $\bar{e}_1$~[GeV$^{-3}$]& - & - & 12.6\\
 $l_3^r(m)$ & 1.4$\cdot 10^{-3}$ & 1.4$\cdot 10^{-3}$ & - \\
 $l_4^r(m)$ & 3.7$\cdot 10^{-3}$ &  3.7$\cdot 10^{-3}$& - \\
 $\mathring{g}$ & 1.0 & 1.3 & -\\
 $\mathring{m}$ [GeV] & 0.87 & 0.87 & 0.8695\\
 $F$ [GeV] & 0.087 & 0.087 & -\\
 $g_A$ & - & - & 1.275\\
 $F_\pi$ [GeV] & - & - & 0.0927\\
 \MN [GeV] & - & - & 0.939\\
 \hline
\end{tabular}
\caption{\label{tbl:sets} Values of low-energy constants employed in the chiral expansions of the nucleon mass $\MN$ (Set \MN)  and $g_A$ (Sets 1 and 2).}
\end{table}

The chiral expansion of the nucleon mass  to order $M^4$ as given in Ref.~\cite{Hoferichter:2025ubp} is
%$M^5$ of Ref.~\cite{McGovern:2006fm},\footnote{While we follow the notation convention of Ref.~\cite{Schindler:2007dr}, these results are obtained in a different formalism.}
%\begin{multline}
%\label{eq:nuc_mass_exp}
%\MN = \mathring{m} + k_1 M^2 +k_2 M^3 + k_3 M^4 \ln(M/\mu)+k^4 M^4\\
%+k_5 M^5 \ln(M/\mu) + k_6 M^5~.
%\end{multline}

\begin{multline}
\label{eq:nuc_mass_exp}
\MN = \mathring{m}-4 c_1 M^2-\frac{3 g_A^2 M^3}{32 \pi F_\pi^2} +\left(k_1\ln\frac{M}{\MN} + k_2 \right)M^4~,
\end{multline}
%We give the explicit form of these coefficients in \eqref{eq:nucmasscoeff} in the appendix.
%The coefficients appearing in the chiral expansion of the nucleon mass in Eq.~\eqref{eq:nuc_mass_exp} are \cite{McGovern:2006fm}
%\begin{align}
%\label{eq:nucmasscoeff}
%\nonumber
%k_1 &= -4 c_1~,\\
%\nonumber
%k_2 & = - \frac{ 3 \mathring{g}^2}{32 \pi F^2 }~,\\
%\nonumber
%k_3 & = - \frac{3}{32 \pi^2 F^2 \mathring{m}}(\mathring{g}^2 -8 c_1 
%\mathring{m} +c_2 \mathring{m}+ 4 c_3 \mathring{m})~,\\
%\nonumber
%k_4 &= \bar{e}_1-\frac{3}{128 \pi^2 F^2 \mathring{m}}(2 \mathring{g}-c_2 \mathring{m})~,\\
%\nonumber
%k_5 & = \frac{3 \mathring{g}^2}{1024 \pi^3 F^4}(16 \mathring{g}^2 -3)~,\\
%\nonumber
%k_6 & = \frac{3}{256 \pi^3 F^4}
%\left( \mathring{g}^2 +\frac{\pi^2 F^2}{\mathring{m}^2} - 
%8 \pi^2\left( 3 l^r_3 -2 l^r_4\right)\right.\\
%&\left.-\frac{32 \pi^2 F^2}{\mathring{g}}\left(2 d^r_{16} -d^r_{18}\right)\right)~.
%\end{align}
%The coefficients appearing in the chiral expansion of the nucleon mass in Eq.~\eqref{eq:nuc_mass_exp} are
with the coefficients
\begin{align}
    \label{eq:nucmasscoeff}
    k_1&= -\frac{3}{32\pi^2 F_\pi^2\MN}\left(g_A^2+\MN\left(-8c_1+c_2+4c_3\right)\right)~,\\
    k_2& =\bar{e}_1 -\frac{3}{128\pi^2 F_\pi^2\MN}\left( 2 g_A^2 -\MN c_2\right)~.
\end{align}
The use of the physical values of the pion decay constant $F_{\pi}$, the nucleon mass \MN and the axial-vector coupling $g_A$ in these expressions instead of the chiral limit values generates differences at order $M^5$, so beyond the order to which we are working. While  \MN has been calculated to order $M^5$ \cite{McGovern:1998tm,McGovern:2006fm,Liang:2025cjd} and order $M^6$ \cite{Schindler:2006ha,Schindler:2007dr,Conrad:2024phd} in various renormalization schemes, using these expressions would require different determinations of the LECs that are currently not available.

\begin{figure}[t]
    \centerline{
    \includegraphics[width=1\columnwidth]{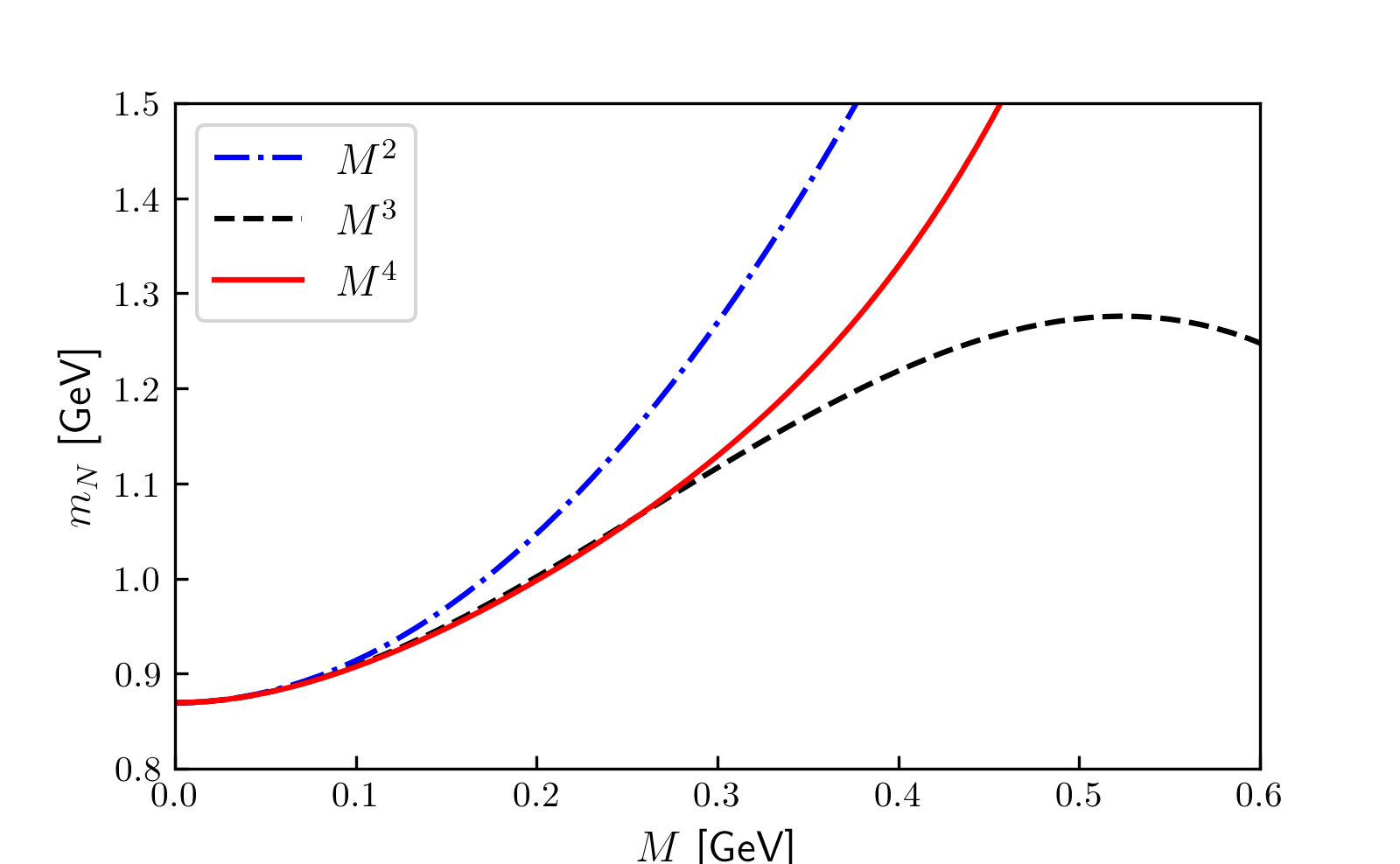}}
    \caption{Pion mass dependence of the nucleon mass \MN using ``Set \MN" given in Table~\ref{tbl:sets}.}
    \label{fig:mn-pi-dep}
\end{figure}

Figure~\ref{fig:mn-pi-dep} shows the pion mass dependence of \MN obtained in this expansion. This result will serve as our data for the Bayesian analysis of the order-by-order convergence of this expansion and enable us to infer the $\chi$PT breakdown scale for \MN. We will perform the same analysis for $g_A$, which we discuss next.

The chiral expansion of the axial-vector coupling constant to order $M^4$ was recently derived in Ref.~\cite{Bernard:2025gto}\footnote{We thank the authors of Ref.~\cite{Bernard:2025gto} for their help in identifying some typos in the original expressions for the expansion coefficients; these are corrected in the following equations.},
\begin{multline}
\label{eq:ga-extrapolation}
g_A = \mathring{g}\Biggl[1+
\left(\frac{\alpha_2}{(4\pi F)^2}\ln\left(\frac{M}{\mu}\right) +\beta_2\right)M^2 
+\alpha_3 M^3\\
+\frac{1}{\mathring{g}}\left( \frac{\alpha_4}{(4\pi F)^4}\ln^2\left(\frac{M}{\mu}\right)\right.\\
\left.+\frac{\gamma_4}{(4\pi F)^4}\ln\left(\frac{M}{\mu}\right)+\beta_4 + \mathring{g} C
\right)M^4
\Biggr]~.
\end{multline}
%[[How about we write]]
%\begin{multline}
%\label{eq:ga-extrapolation}
%g_A = \mathring{g}\Biggl[1+
%\left(\alpha_2\ln\left(\frac{M}{\mu}\right) +\tilde{\beta}_2\right)
%\left(\frac{M}{\tilde{F}}\right)^2
%+\tilde{\alpha}_3 \left(\frac{M}{\tilde{F}}\right)^3\\
%+\left( \alpha_4\ln^2\left(\frac{M}{\mu}\right)
%+\gamma_4\ln\left(\frac{M}{\mu}\right)+\tilde{\beta}_4
%right)\left(\frac{M}{\tilde{F}}\right)^4
%\Biggr]~.
%\end{multline}
%{\bf (MS - changed the first log to frac. Should we add a $+O(M^5)$? If not, need to say in the text to which order we are working)}
The parameters $\alpha_2$ and $\beta_2$ arising from the one-loop calculation were first given in Ref.~\cite{Bernard:1992qa},
\begin{align}
\alpha_2 &= -2 - 4 \mathring{g}^2\\
\beta_2 &= \frac{4}{\mathring{g}} 
%\left(
d_{16}^r(\mu) 
%-2 \mathring{g}d_{28}^r(\mu)
%\right)
-\frac{\mathring{g}^2}{(4 \pi F)^2}~.
\end{align}

%\textcolor{red}{[[MS,LP]]}
%{\bf (MS - with the $beta$ functions given by Bernard, there should not be a $d_{28}$. Its running is included the beta function that Bernard et al use for 16)}

%The definition of $\beta_2$ agrees with Ref.~\cite{Bernard:2006te}, however, the parameter is missing in the definition given in Ref.~\cite{Bernard:2025gto}. 
%The parameters $d_{16}^r$ and $d_{28}^r$ are scale dependent and are related to the scale independent parameters in Tbl.~\ref{tbl:sets} via
%\begin{align}
%d_{16}^r &= \bar{d}_{16}+\frac{\mathring{g}(4-\mathring{g}^2)}
%{8(4 \pi F)^2} \ln(M_\pi^{\rm (phys)}/\mu)~,\\
%d_{28}^r &= \bar{d}_{28} - \frac{9 \mathring{g}^2}{16(4\pi F)^2}
%\ln
%(M_\pi^{\rm phys}/\mu)~.
%\end{align}

%We think, we corrected above a typo in the relation between $d_{28}^r$ and $\bar{d}_{28}$ that is given in \cite{Bernard:2006te}.

The contribution cubic in the pion mass is known to give a large correction and was first calculated in Ref.~\cite{Kambor:1998pi},
\begin{equation}
    \alpha_3 = \frac{1}{24\pi F^2 \mathring{m}}
    (3 + 3 \mathring{g}^2 -4 \mathring{m} c_3+ 8 \mathring{m} c_4)~.
\end{equation}
%\textcolor{red}{[[MS,LP]]}
The expressions for the coefficients of the $M^4$ contribution are given in Ref.~\cite{Bernard:2025gto}.
The coefficient for the leading non-analytic term is 
\begin{align}
    \alpha_4 &= -\frac{7}{3}\mathring{g}(1-\mathring{g}^2)+16 \mathring{g}^5~.
\end{align}
%\textcolor{red}{[[MS, LP]]}
The parameters $\beta_4$ and $\gamma_4$ can be arranged according to the power of $\mathring{g}$ they include,
\begin{align}
    \gamma_4 = &\sum_{i = 0}^5 \mathring{g}^i \gamma_4^{(i)}~,\\
     \beta_4 = &\sum_{i = 0}^5 \mathring{g}^i \beta_4^{(i)}~,
\end{align}
with $\gamma_4^{(4)} =\beta_4^{(4)}=0$. The remaining terms are given by \cite{Bernard:2025gto}
\begin{align}
    \label{eq:beta_i}
    \nonumber
    (4 \pi F)^4 \beta_4^{(0)} &= -4\pi^2 F^2\left( 2 d_{10}^r +4 d_{11}^r
    +3 d_{12}^r +d_{13}^r\right)~,\\
        \nonumber
    (4 \pi F)^4 \beta_4^{(1)} &= -8\pi^2 F^2 \left(d_{14}^r+2(d_1^r+d_2^r) \right) -\frac{\pi^2}{3}\\
    \nonumber
    &\hspace{-17mm}-32\pi^2 l_3^r +\frac{3575}{864}+\frac{1}{2}\psi^{(1)}_{2/3}
    \nonumber
    +\frac{16\pi^2 F^2}{\mathring{m}}( c_2 + 4c_4 +\frac{1}{\mathring{m}})~,\\
           \nonumber
    (4 \pi F)^4 \beta_4^{(2)} &=
    -64\pi^2 F^2\left( 3d_{16}^r - d_{18}^r\right)~,\\
            \nonumber
    (4 \pi F)^4 \beta_4^{(3)} &= 32\pi^2 (-3 l_3^r +l_4^r)
    -\frac{\pi^2}{27}(61+48\log 3)\\
    \nonumber
    &\: -\frac{335}{432}-\frac{1}{6}\psi^{(1)}_{2/3}
    +\frac{32 \pi^2 F^2}{\mathring{m}^2}~,\\
    (4 \pi F)^4 \beta_4^{(5)} &=\frac{41}{36}+\frac{7}{3}\pi^2~,
\end{align}
%\textcolor{red}{[[MS, LP]]}
where $\psi^{(1)}_{2/3}$ denotes the first derivative of the digamma function evaluated at $\textstyle{\frac{2}{3}}$, and 
\begin{align}
\label{eq:gamma4i}
\nonumber
\gamma_4^{(0)} &=16\pi^2 F^2(-20 d_{16}^3+8d_{18}^r+14 d_{10}^r\\
\nonumber
&\qquad+8d_{11}^r+3d_{12}^r + d_{13}^r)~,\\
\nonumber
\gamma_4^{(1)} &= -32\pi^2 F^2(d_{14}^r-2(d_1^r+d_2^r)) -64\pi^2(l_3^r-l_4^r)
\\
\nonumber
&\:-\frac{389}{36}-\frac{64 \pi^2 F^2}{\mathring{m}}\left( c_2+c_3-c_4 - \frac{1}{2\mathring{m}} \right),\\
\nonumber
\gamma_4^{(2)} &= 256\pi^2F^2(d_{18}^r- 3d _{16}^r)~,\\
\nonumber
\gamma_4^{(3)} &= -128\pi^2(l_3^r-l_4^r)+\frac{1}{9}(13-16\pi^2)
+\frac{48\pi^2 F^2}{\mathring{m}^2}~,\\
\gamma_4^{(5)} &=\frac{11}{3}~.
\end{align}
The constant $C$ denotes a linear combination of LECs from the fifth-order pion-nucleon Lagrangian. Its value is currently not determined, with Ref.~\cite{Bernard:2025gto} exploring its impact by considering two non-zero values. We set $C=0$ in our analysis, but have checked that non-zero values such as those considered in Ref.~\cite{Bernard:2025gto} do not significantly impact our findings.

\begin{figure}[t]
    \centerline{
    \includegraphics[width=1\columnwidth]{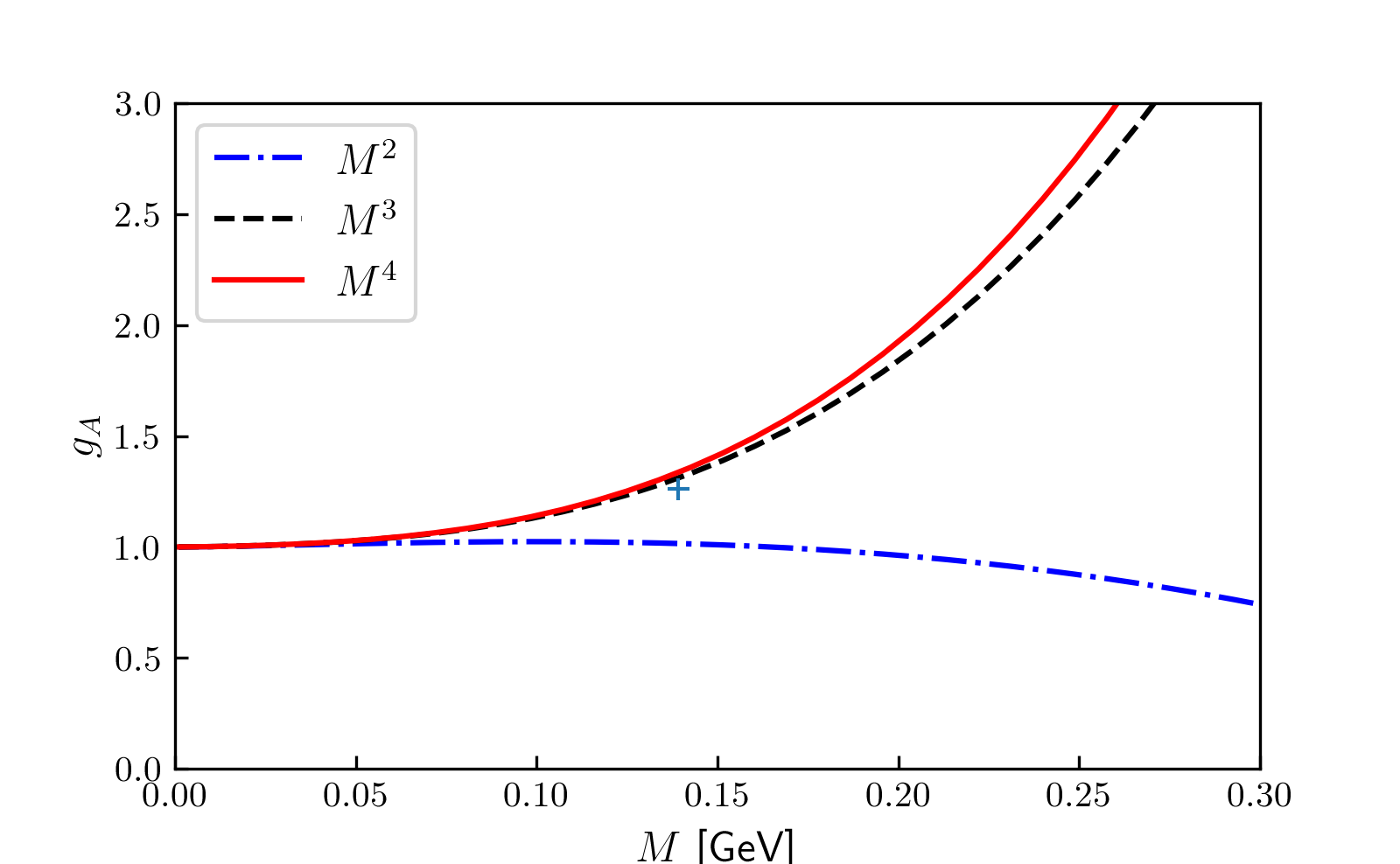}}
    \caption{Pion mass dependence of the axial-vector coupling constant $g_A$ for ``Set 1"  given in Ref.~\cite{Bernard:2025gto} of LECs and $\mathring{g}=1$. For the unknown LEC, we use $C=0$. }
    \label{fig:ga-pi-dep}
\end{figure}

The pion mass dependence of $g_A$ obtained in this expansion with LEC values from ``Set 1" is shown in Fig.~\ref{fig:ga-pi-dep}.

%%%%%
%\subsection{Chiral extrapolations with explicit $\Delta$s}
%The expression for the expansion of the axial coupling constant is
%given in Ref.~\cite{Siemens:2020vop}.
%\begin{equation}
%    g_A = 
%\end{equation}

\section{Analysis and results}
\label{sec:results}
We seek to infer the posterior probability densities $\prob(\Mhi|\bm{y}, I)$ for the breakdown scale \Mhi of $\chi$PT conditioned on a set of order-by-order predictions for the single-nucleon observables \MN and $g_A$, respectively. The data vector $\bm{y}$ contains the chiral expansion for either $g_A$ or \MN, truncated at successive orders, as described in Sec.~\ref{sec:chiralexpansions}. The inference is also conditioned on assumptions, denoted $I$, that are outlined in Sec.~\ref{sec:chiralexpansions} and below. Intuitively, the posterior for \Mhi reflects how well the chiral expansion converges across successive orders and encodes the extent to which the observed convergence pattern is consistent with the natural scales of the $\chi$PT power counting, as reflected in the priors we place.

We use the point-wise method presented in~\citet{Melendez:2019izc} and thus first use Bayes' theorem to express the sought posterior as proportional to a likelihood times a prior
\begin{equation}
\prob(\Mhi|\bm{y},I) \propto \prob(\bm{y}|\Mhi,I)\cdot
\prob(\Mhi|I).  \label{eq:posterior}
\end{equation} 

\begin{figure}[t]
    \centerline{
  \includegraphics[width=0.9\columnwidth,clip=true]{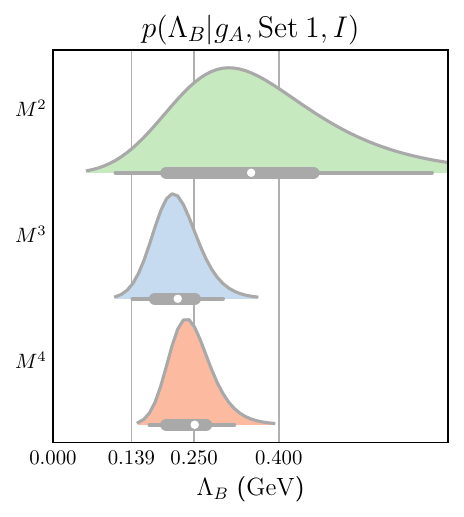}}
  \centerline{
    \includegraphics[width=0.9 \columnwidth,clip=true]{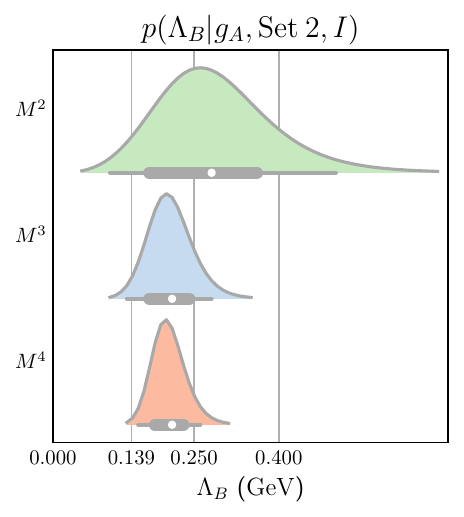}}
    \caption{Posteriors for the breakdown scale \Mhi obtained from the analysis of the order-by-order expansion of the axial-vector coupling constant
    $g_A$ and using the values from Set 1 (top panel) and Set 2 (bottom panel) in Table~\ref{tbl:sets} for the LECs. }
      \label{fig:gA-break}
\end{figure}

\begin{figure}[t]
    \centerline{
    \includegraphics[width=0.9\columnwidth,clip=true]{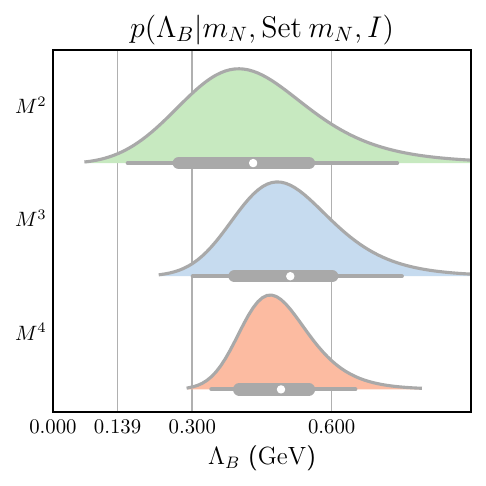}}
    \caption{Posteriors for the breakdown scale \Mhi obtained from the analysis of the order-by-order expansion of nucleon mass \MN and using the values from ``Set \MN" in Table~\ref{tbl:sets} for the LECs.}
    \label{fig:nm-break}
\end{figure}

For the likelihood $\prob(\bm{y}|\Mhi,I)$ we assume the $n$-th chiral order prediction for an observable $X\in\{g_A,m_N\}$ to follow a formal expansion~\cite{Furnstahl:2015rha} 
\begin{align} X^{(n)} & =
X_\text{ref} \sum_{i=0}^{n} c_i Q^i~, 
\label{eq:eft_expansion}
\end{align} with
dimensionless expansion parameter $Q = M/\Mhi$ and coefficients $c_i$\footnote{Note that these $c_i$ coefficients are distinct and not the same as the LECs entering the chiral expansions.}. The breakdown scale  for the observable $X$ corresponds to the pion mass at which all terms in its $\chi$PT expansion give contributions of the same size. Therefore, once $M$ actually reaches \Mhi we are well beyond the point at which $\chi$PT is a useful calculational tool for $X$. We assume natural-sized coefficients by
assigning a normal prior $c_i \overset{\text{iid}}{\sim} \mathcal{N}(0, \bar{c}^2)$ with a conjugate inverse-chi-squared $\chi^{-2}$ hyperprior for the variance $\bar{c}^2$, i.e., we have $\bar{c}^2 \sim \chi^{-2}(\nu_0 = 2, \tau_0^2 =1)$. This choice places about 67\% of the probability for $\bar{c}^2$ within the natural range $ [1/3, 3]$.  For both observables, we set the reference scale $X_\text{ref}$ equal to the leading-order contribution, so that $c_0=1$. We then quantify the expansion coefficients $c_i$ in Eq.~\eqref{eq:eft_expansion} from the order-by-order calculations of $X$ at $M=[9.85,301.84]$~MeV. Here, we selected expansion data at two $M$-values to avoid over-confident posteriors by remaining consistent with the assumption of independent data\footnote{One might be concerned that this interval includes $M$ values that are potentially larger than \Mhi. However, this does not invalidate the use of Eq.~(\ref{eq:posterior}) to calculate the \Mhi posterior.}.
For the prior $p(\Mhi|I)$ we assume a log-uniform
distribution with compact support on the interval $[\Mpi^{(\rm phys) }/40,40\Mpi^{(\rm phys) }]$. 
From this we infer posterior probability distributions for the $\chi$PT breakdown scales for $g_A$ and \MN as shown in Figs.~\ref{fig:gA-break}-\ref{fig:nm-break}, respectively. The order-by-order posteriors for $\Mhi$ converge steadily, for both observables. 
The results for median values and 68\% degree of belief intervals are summarized in Table~\ref{tbl:breakdownscales}.

We checked that our Bayesian analysis is robust with respect to variations in the employed priors and to different choices of point-wise $M$-values used for extracting the order-by-order expansion data. As we increase the number of soft-scale evaluation points, the posterior modes of the breakdown scales remain unchanged, while the corresponding distributions become narrower. This is not a true reduction in posterior uncertainty since our analysis does not account for the finite correlation length of the EFT expansion coefficients $c_i$. 
 
\begin{table}[t]
\begin{tabular}{ |p{1.cm}||p{2cm}|p{2cm}| p{2.4cm} |}
 \hline
 Order & $\Lambda_B(g_A$;Set 1$)$ & $\Lambda_B(g_A$;Set 2$)$ & $\Lambda_B(m_N$;Set $m_N)$\\
\hline
$M^2$  & 351 (201,461) & 281 (171,361) & 431 (271,551)\\
$M^3$  & 221 (181,251) & 211 (171,241) & 511 (391,601)\\
$M^4$  & 251 (201,271) & 211 (181,231) & 491 (401,551)\\
\hline
\end{tabular}
\caption{\label{tbl:breakdownscales}
Median values and 68\% degree of belief intervals (highest posterior density) for the breakdown scales, obtained at different orders in the chiral expansions for $g_A$ and \MN. All values in MeV.}
\end{table}

\section{Summary and discussion}
\label{sec:summary}
We have used Bayesian methods to infer a breakdown scale of approximately $230$~MeV when analyzing the $\chi$PT expansion of the axial-vector coupling $g_A$ and 490~MeV when considering the expansion of the nucleon mass, \MN. We note that these values for \Mhi were derived using point estimates of the LECs that appear in the $\chi$PT expansion of \MN and $g_A$. A complete Bayesian analysis of the convergence of the EFT would estimate the LECs from data in the presence of truncation errors and then consistently determine \Mhi as part of that inference process. Refs.~\cite{Wesolowski:2021cni,Bub:2024gyz} provide examples of such analyses for the case of $\chi$EFT applied to few-nucleon systems. 

While it has been known for some time that the convergence of baryon $\chi$PT for $g_A$ is worse than that for \MN, our Bayesian analysis quantifies the strength of this inference. 
%Our analysis quantifies therefore for the first time the extent to which these two observables are sensitive to different scales: 
The much lower number found for \Mhi in the case of the expansion of $g_A$ 
appears to be related to the nucleon-$\Delta$ mass splitting. %, we do not observe the same for the breakdown scale inferred from the nucleon mass expansion. 
Meanwhile, the breakdown scale inferred in the case of \MN
%matches rather the scale associated 
is closer to the mass scale of  light mesonic degrees of freedom not  explicitly included in $\chi$PT.

%While it has long been argued that the presence of the $\Delta$ excitation slows the convergence of the chiral expansion of the axial coupling constant, our analysis demonstrates for the first time that the inferred breakdown scale indeed lies indeed close to the nucleon–$\Delta$ mass splitting.

%Our and previous results demonstrate 

This emphasizes
that the breakdown scale inferred from an effective field theory expansion of an observable will typically
depend on the observable(s) used for inference. We note that the breakdown scale of nucleon-nucleon scattering in chiral effective theory has been inferred to be higher than both breakdown scales found here, at approximately 600~MeV~\cite{Melendez:2017phj,Wesolowski:2021cni,Drischler:2020hwi,Millican:2024yuz,McClung:2025rtc}. Even though there is evidence that the breakdown scale in nucleon-nucleon scattering is lower in a strictly renormalizable implementation of $\chi$EFT~\cite{Thim:2024yks}, 
nucleon-nucleon scattering does not appear to show strong sensitivity to the presence of the $\Delta$-excitation. 

However, it remains an open question whether there are certain nuclear matrix elements that are influenced by $\Delta$ excitation and so also exhibit a lower \Mhi. For example, the two-nucleon axial current involves the linear combination $c_3 + 2c_4$ of pion-nucleon LECs. These are the same LECs responsible for the large $M^3$ correction in the chiral expansion of $g_A$. However, in the case of the two-nucleon axial current they occur with the same sign, leading to cancellation instead of reinforcement. Thus, ultimately 
the contribution of this current to most nuclear $\beta$-decay matrix elements is of the size expected from the $\chi$EFT counting.

Our results also raise the important question whether the breakdown scale inferred for observables in powers of the pion mass $M$ is the same for observables that are also momentum-dependent and thus also expanded in powers of the momentum ${\bf q}$. This outcome would, for example, have important consequences for $\chi$PT calculations of form factors such as the axial form factor.

\begin{acknowledgments}
We are grateful to J.~Gegelia and U.-G.~Mei\ss ner for answering our questions regarding Ref.~\cite{Bernard:2025gto}, and to V.~Bernard for a detailed comparison of the expressions for the chiral expansion of $g_A$. MRS thanks T.~R.~Richardson for helpful discussions.
This work was supported by the National Science Foundation (Grant Nos.\ PHY-2111426 and PHY-2412612), the  US Department of Energy (Contract No. DE-AC05-00OR22725), the U.S.~Department of Energy, Office of Science, Office of Nuclear Physics (Award Nos.~DE-SC0019647  and DE-FG02-93ER-40756) and by the Swedish Research Council (Grants No. 2020-05127 and No. 2024-04681).
We thank the Institute for Nuclear Theory at the University of Washington for its kind hospitality and stimulating research environment. This research was supported in part by the INT's U.S. Department of Energy grant No. DE-FG02-00ER41132.
\end{acknowledgments}
\bibliography{biblio}

\begin{thebibliography}{52}%
\makeatletter
\providecommand \@ifxundefined [1]{%
 \@ifx{#1\undefined}
}%
\providecommand \@ifnum [1]{%
 \ifnum #1\expandafter \@firstoftwo
 \else \expandafter \@secondoftwo
 \fi
}%
\providecommand \@ifx [1]{%
 \ifx #1\expandafter \@firstoftwo
 \else \expandafter \@secondoftwo
 \fi
}%
\providecommand \natexlab [1]{#1}%
\providecommand \enquote  [1]{``#1''}%
\providecommand \bibnamefont  [1]{#1}%
\providecommand \bibfnamefont [1]{#1}%
\providecommand \citenamefont [1]{#1}%
\providecommand \href@noop [0]{\@secondoftwo}%
\providecommand \href [0]{\begingroup \@sanitize@url \@href}%
\providecommand \@href[1]{\@@startlink{#1}\@@href}%
\providecommand \@@href[1]{\endgroup#1\@@endlink}%
\providecommand \@sanitize@url [0]{\catcode `\\12\catcode `\$12\catcode `\&12\catcode `\#12\catcode `\^12\catcode `\_12\catcode `\%12\relax}%
\providecommand \@@startlink[1]{}%
\providecommand \@@endlink[0]{}%
\providecommand \url  [0]{\begingroup\@sanitize@url \@url }%
\providecommand \@url [1]{\endgroup\@href {#1}{\urlprefix }}%
\providecommand \urlprefix  [0]{URL }%
\providecommand \Eprint [0]{\href }%
\providecommand \doibase [0]{http://dx.doi.org/}%
\providecommand \selectlanguage [0]{\@gobble}%
\providecommand \bibinfo  [0]{\@secondoftwo}%
\providecommand \bibfield  [0]{\@secondoftwo}%
\providecommand \translation [1]{[#1]}%
\providecommand \BibitemOpen [0]{}%
\providecommand \bibitemStop [0]{}%
\providecommand \bibitemNoStop [0]{.\EOS\space}%
\providecommand \EOS [0]{\spacefactor3000\relax}%
\providecommand \BibitemShut  [1]{\csname bibitem#1\endcsname}%
\let\auto@bib@innerbib\@empty
%</preamble>
\bibitem [{\citenamefont {Weinberg}(1979)}]{Weinberg:1978kz}%
  \BibitemOpen
  \bibfield  {author} {\bibinfo {author} {\bibfnamefont {S.}~\bibnamefont {Weinberg}},\ }\href {\doibase 10.1016/0378-4371(79)90223-1} {\bibfield  {journal} {\bibinfo  {journal} {Physica A}\ }\textbf {\bibinfo {volume} {96}},\ \bibinfo {pages} {327} (\bibinfo {year} {1979})}\BibitemShut {NoStop}%
\bibitem [{\citenamefont {Gasser}\ and\ \citenamefont {Leutwyler}(1984)}]{Gasser:1983yg}%
  \BibitemOpen
  \bibfield  {author} {\bibinfo {author} {\bibfnamefont {J.}~\bibnamefont {Gasser}}\ and\ \bibinfo {author} {\bibfnamefont {H.}~\bibnamefont {Leutwyler}},\ }\href {\doibase 10.1016/0003-4916(84)90242-2} {\bibfield  {journal} {\bibinfo  {journal} {Annals Phys.}\ }\textbf {\bibinfo {volume} {158}},\ \bibinfo {pages} {142} (\bibinfo {year} {1984})}\BibitemShut {NoStop}%
\bibitem [{\citenamefont {Gasser}\ and\ \citenamefont {Leutwyler}(1985)}]{Gasser:1984gg}%
  \BibitemOpen
  \bibfield  {author} {\bibinfo {author} {\bibfnamefont {J.}~\bibnamefont {Gasser}}\ and\ \bibinfo {author} {\bibfnamefont {H.}~\bibnamefont {Leutwyler}},\ }\href {\doibase 10.1016/0550-3213(85)90492-4} {\bibfield  {journal} {\bibinfo  {journal} {Nucl. Phys. B}\ }\textbf {\bibinfo {volume} {250}},\ \bibinfo {pages} {465} (\bibinfo {year} {1985})}\BibitemShut {NoStop}%
\bibitem [{\citenamefont {Manohar}\ and\ \citenamefont {Georgi}(1984)}]{Manohar:1983md}%
  \BibitemOpen
  \bibfield  {author} {\bibinfo {author} {\bibfnamefont {A.}~\bibnamefont {Manohar}}\ and\ \bibinfo {author} {\bibfnamefont {H.}~\bibnamefont {Georgi}},\ }\href {\doibase 10.1016/0550-3213(84)90231-1} {\bibfield  {journal} {\bibinfo  {journal} {Nucl. Phys. B}\ }\textbf {\bibinfo {volume} {234}},\ \bibinfo {pages} {189} (\bibinfo {year} {1984})}\BibitemShut {NoStop}%
\bibitem [{\citenamefont {Gasser}\ \emph {et~al.}(1988)\citenamefont {Gasser}, \citenamefont {Sainio},\ and\ \citenamefont {\v{S}varc}}]{Gasser:1987rb}%
  \BibitemOpen
  \bibfield  {author} {\bibinfo {author} {\bibfnamefont {J.}~\bibnamefont {Gasser}}, \bibinfo {author} {\bibfnamefont {M.~E.}\ \bibnamefont {Sainio}}, \ and\ \bibinfo {author} {\bibfnamefont {A.}~\bibnamefont {\v{S}varc}},\ }\href {\doibase 10.1016/0550-3213(88)90108-3} {\bibfield  {journal} {\bibinfo  {journal} {Nucl. Phys. B}\ }\textbf {\bibinfo {volume} {307}},\ \bibinfo {pages} {779} (\bibinfo {year} {1988})}\BibitemShut {NoStop}%
\bibitem [{\citenamefont {Jenkins}\ and\ \citenamefont {Manohar}(1991{\natexlab{a}})}]{Jenkins:1990jv}%
  \BibitemOpen
  \bibfield  {author} {\bibinfo {author} {\bibfnamefont {E.~E.}\ \bibnamefont {Jenkins}}\ and\ \bibinfo {author} {\bibfnamefont {A.~V.}\ \bibnamefont {Manohar}},\ }\href {\doibase 10.1016/0370-2693(91)90266-S} {\bibfield  {journal} {\bibinfo  {journal} {Phys. Lett. B}\ }\textbf {\bibinfo {volume} {255}},\ \bibinfo {pages} {558} (\bibinfo {year} {1991}{\natexlab{a}})}\BibitemShut {NoStop}%
\bibitem [{\citenamefont {Bernard}\ \emph {et~al.}(1992)\citenamefont {Bernard}, \citenamefont {Kaiser}, \citenamefont {Kambor},\ and\ \citenamefont {Meissner}}]{Bernard:1992qa}%
  \BibitemOpen
  \bibfield  {author} {\bibinfo {author} {\bibfnamefont {V.}~\bibnamefont {Bernard}}, \bibinfo {author} {\bibfnamefont {N.}~\bibnamefont {Kaiser}}, \bibinfo {author} {\bibfnamefont {J.}~\bibnamefont {Kambor}}, \ and\ \bibinfo {author} {\bibfnamefont {U.~G.}\ \bibnamefont {Meissner}},\ }\href {\doibase 10.1016/0550-3213(92)90615-I} {\bibfield  {journal} {\bibinfo  {journal} {Nucl. Phys. B}\ }\textbf {\bibinfo {volume} {388}},\ \bibinfo {pages} {315} (\bibinfo {year} {1992})}\BibitemShut {NoStop}%
\bibitem [{\citenamefont {Becher}\ and\ \citenamefont {Leutwyler}(1999)}]{Becher:1999he}%
  \BibitemOpen
  \bibfield  {author} {\bibinfo {author} {\bibfnamefont {T.}~\bibnamefont {Becher}}\ and\ \bibinfo {author} {\bibfnamefont {H.}~\bibnamefont {Leutwyler}},\ }\href {\doibase 10.1007/PL00021673} {\bibfield  {journal} {\bibinfo  {journal} {Eur. Phys. J.}\ }\textbf {\bibinfo {volume} {9}},\ \bibinfo {pages} {643} (\bibinfo {year} {1999})},\ \Eprint {http://arxiv.org/abs/hep-ph/9901384} {arXiv:hep-ph/9901384} \BibitemShut {NoStop}%
\bibitem [{\citenamefont {Gegelia}\ and\ \citenamefont {Japaridze}(1999)}]{Gegelia:1999gf}%
  \BibitemOpen
  \bibfield  {author} {\bibinfo {author} {\bibfnamefont {J.}~\bibnamefont {Gegelia}}\ and\ \bibinfo {author} {\bibfnamefont {G.}~\bibnamefont {Japaridze}},\ }\href {\doibase 10.1103/PhysRevD.60.114038} {\bibfield  {journal} {\bibinfo  {journal} {Phys. Rev. D}\ }\textbf {\bibinfo {volume} {60}},\ \bibinfo {pages} {114038} (\bibinfo {year} {1999})},\ \Eprint {http://arxiv.org/abs/hep-ph/9908377} {arXiv:hep-ph/9908377} \BibitemShut {NoStop}%
\bibitem [{\citenamefont {Fuchs}\ \emph {et~al.}(2003)\citenamefont {Fuchs}, \citenamefont {Gegelia}, \citenamefont {Japaridze},\ and\ \citenamefont {Scherer}}]{Fuchs:2003qc}%
  \BibitemOpen
  \bibfield  {author} {\bibinfo {author} {\bibfnamefont {T.}~\bibnamefont {Fuchs}}, \bibinfo {author} {\bibfnamefont {J.}~\bibnamefont {Gegelia}}, \bibinfo {author} {\bibfnamefont {G.}~\bibnamefont {Japaridze}}, \ and\ \bibinfo {author} {\bibfnamefont {S.}~\bibnamefont {Scherer}},\ }\href {\doibase 10.1103/PhysRevD.68.056005} {\bibfield  {journal} {\bibinfo  {journal} {Phys. Rev. D}\ }\textbf {\bibinfo {volume} {68}},\ \bibinfo {pages} {056005} (\bibinfo {year} {2003})},\ \Eprint {http://arxiv.org/abs/hep-ph/0302117} {arXiv:hep-ph/0302117} \BibitemShut {NoStop}%
\bibitem [{\citenamefont {Bernard}(2008)}]{Bernard:2007zu}%
  \BibitemOpen
  \bibfield  {author} {\bibinfo {author} {\bibfnamefont {V.}~\bibnamefont {Bernard}},\ }\href {\doibase 10.1016/j.ppnp.2007.07.001} {\bibfield  {journal} {\bibinfo  {journal} {Prog. Part. Nucl. Phys.}\ }\textbf {\bibinfo {volume} {60}},\ \bibinfo {pages} {82} (\bibinfo {year} {2008})},\ \Eprint {http://arxiv.org/abs/0706.0312} {arXiv:0706.0312 [hep-ph]} \BibitemShut {NoStop}%
\bibitem [{\citenamefont {Scherer}\ and\ \citenamefont {Schindler}(2012)}]{Scherer:2012xha}%
  \BibitemOpen
  \bibfield  {author} {\bibinfo {author} {\bibfnamefont {S.}~\bibnamefont {Scherer}}\ and\ \bibinfo {author} {\bibfnamefont {M.~R.}\ \bibnamefont {Schindler}},\ }\href {\doibase 10.1007/978-3-642-19254-8} {\emph {\bibinfo {title} {{A Primer for Chiral Perturbation Theory}}}},\ Vol.\ \bibinfo {volume} {830}\ (\bibinfo {year} {2012})\BibitemShut {NoStop}%
\bibitem [{\citenamefont {Bernard}\ \emph {et~al.}(1991)\citenamefont {Bernard}, \citenamefont {Kaiser},\ and\ \citenamefont {Meissner}}]{Bernard:1991rq}%
  \BibitemOpen
  \bibfield  {author} {\bibinfo {author} {\bibfnamefont {V.}~\bibnamefont {Bernard}}, \bibinfo {author} {\bibfnamefont {N.}~\bibnamefont {Kaiser}}, \ and\ \bibinfo {author} {\bibfnamefont {U.~G.}\ \bibnamefont {Meissner}},\ }\href {\doibase 10.1103/PhysRevLett.67.1515} {\bibfield  {journal} {\bibinfo  {journal} {Phys. Rev. Lett.}\ }\textbf {\bibinfo {volume} {67}},\ \bibinfo {pages} {1515} (\bibinfo {year} {1991})}\BibitemShut {NoStop}%
\bibitem [{\citenamefont {Bernard}\ \emph {et~al.}(1995)\citenamefont {Bernard}, \citenamefont {Kaiser},\ and\ \citenamefont {Meissner}}]{Bernard:1995dp}%
  \BibitemOpen
  \bibfield  {author} {\bibinfo {author} {\bibfnamefont {V.}~\bibnamefont {Bernard}}, \bibinfo {author} {\bibfnamefont {N.}~\bibnamefont {Kaiser}}, \ and\ \bibinfo {author} {\bibfnamefont {U.-G.}\ \bibnamefont {Meissner}},\ }\href {\doibase 10.1142/S0218301395000092} {\bibfield  {journal} {\bibinfo  {journal} {Int. J. Mod. Phys. E}\ }\textbf {\bibinfo {volume} {4}},\ \bibinfo {pages} {193} (\bibinfo {year} {1995})},\ \Eprint {http://arxiv.org/abs/hep-ph/9501384} {arXiv:hep-ph/9501384} \BibitemShut {NoStop}%
\bibitem [{\citenamefont {Jenkins}\ and\ \citenamefont {Manohar}(1991{\natexlab{b}})}]{Jenkins:1991ne}%
  \BibitemOpen
  \bibfield  {author} {\bibinfo {author} {\bibfnamefont {E.~E.}\ \bibnamefont {Jenkins}}\ and\ \bibinfo {author} {\bibfnamefont {A.~V.}\ \bibnamefont {Manohar}},\ }in\ \href@noop {} {\emph {\bibinfo {booktitle} {{Workshop on Effective Field Theories of the Standard Model}}}}\ (\bibinfo {year} {1991})\BibitemShut {NoStop}%
\bibitem [{\citenamefont {Hemmert}\ \emph {et~al.}(1997)\citenamefont {Hemmert}, \citenamefont {Holstein},\ and\ \citenamefont {Kambor}}]{Hemmert:1996xg}%
  \BibitemOpen
  \bibfield  {author} {\bibinfo {author} {\bibfnamefont {T.~R.}\ \bibnamefont {Hemmert}}, \bibinfo {author} {\bibfnamefont {B.~R.}\ \bibnamefont {Holstein}}, \ and\ \bibinfo {author} {\bibfnamefont {J.}~\bibnamefont {Kambor}},\ }\href {\doibase 10.1016/S0370-2693(97)00049-X} {\bibfield  {journal} {\bibinfo  {journal} {Phys. Lett. B}\ }\textbf {\bibinfo {volume} {395}},\ \bibinfo {pages} {89} (\bibinfo {year} {1997})},\ \Eprint {http://arxiv.org/abs/hep-ph/9606456} {arXiv:hep-ph/9606456} \BibitemShut {NoStop}%
\bibitem [{\citenamefont {Hemmert}\ \emph {et~al.}(1998)\citenamefont {Hemmert}, \citenamefont {Holstein},\ and\ \citenamefont {Kambor}}]{Hemmert:1997ye}%
  \BibitemOpen
  \bibfield  {author} {\bibinfo {author} {\bibfnamefont {T.~R.}\ \bibnamefont {Hemmert}}, \bibinfo {author} {\bibfnamefont {B.~R.}\ \bibnamefont {Holstein}}, \ and\ \bibinfo {author} {\bibfnamefont {J.}~\bibnamefont {Kambor}},\ }\href {\doibase 10.1088/0954-3899/24/10/003} {\bibfield  {journal} {\bibinfo  {journal} {J. Phys. G}\ }\textbf {\bibinfo {volume} {24}},\ \bibinfo {pages} {1831} (\bibinfo {year} {1998})},\ \Eprint {http://arxiv.org/abs/hep-ph/9712496} {arXiv:hep-ph/9712496} \BibitemShut {NoStop}%
\bibitem [{\citenamefont {Jenkins}\ and\ \citenamefont {Manohar}(1991{\natexlab{c}})}]{Jenkins:1991es}%
  \BibitemOpen
  \bibfield  {author} {\bibinfo {author} {\bibfnamefont {E.~E.}\ \bibnamefont {Jenkins}}\ and\ \bibinfo {author} {\bibfnamefont {A.~V.}\ \bibnamefont {Manohar}},\ }\href {\doibase 10.1016/0370-2693(91)90840-M} {\bibfield  {journal} {\bibinfo  {journal} {Phys. Lett. B}\ }\textbf {\bibinfo {volume} {259}},\ \bibinfo {pages} {353} (\bibinfo {year} {1991}{\natexlab{c}})}\BibitemShut {NoStop}%
\bibitem [{\citenamefont {Luty}\ and\ \citenamefont {White}(1993)}]{Luty:1993gi}%
  \BibitemOpen
  \bibfield  {author} {\bibinfo {author} {\bibfnamefont {M.~A.}\ \bibnamefont {Luty}}\ and\ \bibinfo {author} {\bibfnamefont {M.~J.}\ \bibnamefont {White}},\ }\href {\doibase 10.1016/0370-2693(93)90812-V} {\bibfield  {journal} {\bibinfo  {journal} {Phys. Lett. B}\ }\textbf {\bibinfo {volume} {319}},\ \bibinfo {pages} {261} (\bibinfo {year} {1993})},\ \Eprint {http://arxiv.org/abs/hep-ph/9305203} {arXiv:hep-ph/9305203} \BibitemShut {NoStop}%
\bibitem [{\citenamefont {Flores-Mendieta}\ \emph {et~al.}(2000)\citenamefont {Flores-Mendieta}, \citenamefont {Hofmann}, \citenamefont {Jenkins},\ and\ \citenamefont {Manohar}}]{Flores-Mendieta:2000ljq}%
  \BibitemOpen
  \bibfield  {author} {\bibinfo {author} {\bibfnamefont {R.}~\bibnamefont {Flores-Mendieta}}, \bibinfo {author} {\bibfnamefont {C.~P.}\ \bibnamefont {Hofmann}}, \bibinfo {author} {\bibfnamefont {E.~E.}\ \bibnamefont {Jenkins}}, \ and\ \bibinfo {author} {\bibfnamefont {A.~V.}\ \bibnamefont {Manohar}},\ }\href {\doibase 10.1103/PhysRevD.62.034001} {\bibfield  {journal} {\bibinfo  {journal} {Phys. Rev. D}\ }\textbf {\bibinfo {volume} {62}},\ \bibinfo {pages} {034001} (\bibinfo {year} {2000})},\ \Eprint {http://arxiv.org/abs/hep-ph/0001218} {arXiv:hep-ph/0001218} \BibitemShut {NoStop}%
\bibitem [{\citenamefont {Hemmert}\ \emph {et~al.}(2003)\citenamefont {Hemmert}, \citenamefont {Procura},\ and\ \citenamefont {Weise}}]{Hemmert:2003cb}%
  \BibitemOpen
  \bibfield  {author} {\bibinfo {author} {\bibfnamefont {T.~R.}\ \bibnamefont {Hemmert}}, \bibinfo {author} {\bibfnamefont {M.}~\bibnamefont {Procura}}, \ and\ \bibinfo {author} {\bibfnamefont {W.}~\bibnamefont {Weise}},\ }\href {\doibase 10.1103/PhysRevD.68.075009} {\bibfield  {journal} {\bibinfo  {journal} {Phys. Rev. D}\ }\textbf {\bibinfo {volume} {68}},\ \bibinfo {pages} {075009} (\bibinfo {year} {2003})},\ \Eprint {http://arxiv.org/abs/hep-lat/0303002} {arXiv:hep-lat/0303002} \BibitemShut {NoStop}%
\bibitem [{\citenamefont {Calle~Cordon}\ and\ \citenamefont {Goity}(2013)}]{CalleCordon:2012xz}%
  \BibitemOpen
  \bibfield  {author} {\bibinfo {author} {\bibfnamefont {A.}~\bibnamefont {Calle~Cordon}}\ and\ \bibinfo {author} {\bibfnamefont {J.~L.}\ \bibnamefont {Goity}},\ }\href {\doibase 10.1103/PhysRevD.87.016019} {\bibfield  {journal} {\bibinfo  {journal} {Phys. Rev. D}\ }\textbf {\bibinfo {volume} {87}},\ \bibinfo {pages} {016019} (\bibinfo {year} {2013})},\ \Eprint {http://arxiv.org/abs/1210.2364} {arXiv:1210.2364 [nucl-th]} \BibitemShut {NoStop}%
\bibitem [{\citenamefont {Hall}\ \emph {et~al.}(2025)\citenamefont {Hall} \emph {et~al.}}]{Hall:2025ytt}%
  \BibitemOpen
  \bibfield  {author} {\bibinfo {author} {\bibfnamefont {Z.~B.}\ \bibnamefont {Hall}} \emph {et~al.},\ }\href@noop {} {\  (\bibinfo {year} {2025})},\ \Eprint {http://arxiv.org/abs/2503.09891} {arXiv:2503.09891 [hep-lat]} \BibitemShut {NoStop}%
\bibitem [{\citenamefont {Dashen}\ \emph {et~al.}(1994)\citenamefont {Dashen}, \citenamefont {Jenkins},\ and\ \citenamefont {Manohar}}]{Dashen:1993jt}%
  \BibitemOpen
  \bibfield  {author} {\bibinfo {author} {\bibfnamefont {R.~F.}\ \bibnamefont {Dashen}}, \bibinfo {author} {\bibfnamefont {E.~E.}\ \bibnamefont {Jenkins}}, \ and\ \bibinfo {author} {\bibfnamefont {A.~V.}\ \bibnamefont {Manohar}},\ }\href {\doibase 10.1103/PhysRevD.51.2489} {\bibfield  {journal} {\bibinfo  {journal} {Phys. Rev. D}\ }\textbf {\bibinfo {volume} {49}},\ \bibinfo {pages} {4713} (\bibinfo {year} {1994})},\ \bibinfo {note} {[Erratum: Phys.Rev.D 51, 2489 (1995)]},\ \Eprint {http://arxiv.org/abs/hep-ph/9310379} {arXiv:hep-ph/9310379} \BibitemShut {NoStop}%
\bibitem [{\citenamefont {Djukanovic}\ \emph {et~al.}(2006)\citenamefont {Djukanovic}, \citenamefont {Gegelia},\ and\ \citenamefont {Scherer}}]{Djukanovic:2006xc}%
  \BibitemOpen
  \bibfield  {author} {\bibinfo {author} {\bibfnamefont {D.}~\bibnamefont {Djukanovic}}, \bibinfo {author} {\bibfnamefont {J.}~\bibnamefont {Gegelia}}, \ and\ \bibinfo {author} {\bibfnamefont {S.}~\bibnamefont {Scherer}},\ }\href {\doibase 10.1140/epja/i2006-10096-6} {\bibfield  {journal} {\bibinfo  {journal} {Eur. Phys. J. A}\ }\textbf {\bibinfo {volume} {29}},\ \bibinfo {pages} {337} (\bibinfo {year} {2006})},\ \Eprint {http://arxiv.org/abs/hep-ph/0604164} {arXiv:hep-ph/0604164} \BibitemShut {NoStop}%
\bibitem [{\citenamefont {McGovern}\ and\ \citenamefont {Birse}(2006)}]{McGovern:2006fm}%
  \BibitemOpen
  \bibfield  {author} {\bibinfo {author} {\bibfnamefont {J.~A.}\ \bibnamefont {McGovern}}\ and\ \bibinfo {author} {\bibfnamefont {M.~C.}\ \bibnamefont {Birse}},\ }\href {\doibase 10.1103/PhysRevD.74.097501} {\bibfield  {journal} {\bibinfo  {journal} {Phys. Rev. D}\ }\textbf {\bibinfo {volume} {74}},\ \bibinfo {pages} {097501} (\bibinfo {year} {2006})},\ \Eprint {http://arxiv.org/abs/hep-lat/0608002} {arXiv:hep-lat/0608002} \BibitemShut {NoStop}%
\bibitem [{\citenamefont {Schindler}\ \emph {et~al.}(2007)\citenamefont {Schindler}, \citenamefont {Djukanovic}, \citenamefont {Gegelia},\ and\ \citenamefont {Scherer}}]{Schindler:2006ha}%
  \BibitemOpen
  \bibfield  {author} {\bibinfo {author} {\bibfnamefont {M.~R.}\ \bibnamefont {Schindler}}, \bibinfo {author} {\bibfnamefont {D.}~\bibnamefont {Djukanovic}}, \bibinfo {author} {\bibfnamefont {J.}~\bibnamefont {Gegelia}}, \ and\ \bibinfo {author} {\bibfnamefont {S.}~\bibnamefont {Scherer}},\ }\href {\doibase 10.1016/j.physletb.2007.04.034} {\bibfield  {journal} {\bibinfo  {journal} {Phys. Lett. B}\ }\textbf {\bibinfo {volume} {649}},\ \bibinfo {pages} {390} (\bibinfo {year} {2007})},\ \Eprint {http://arxiv.org/abs/hep-ph/0612164} {arXiv:hep-ph/0612164} \BibitemShut {NoStop}%
\bibitem [{\citenamefont {Bernard}\ and\ \citenamefont {Meissner}(2006)}]{Bernard:2006te}%
  \BibitemOpen
  \bibfield  {author} {\bibinfo {author} {\bibfnamefont {V.}~\bibnamefont {Bernard}}\ and\ \bibinfo {author} {\bibfnamefont {U.-G.}\ \bibnamefont {Meissner}},\ }\href {\doibase 10.1016/j.physletb.2006.06.018} {\bibfield  {journal} {\bibinfo  {journal} {Phys. Lett. B}\ }\textbf {\bibinfo {volume} {639}},\ \bibinfo {pages} {278} (\bibinfo {year} {2006})},\ \Eprint {http://arxiv.org/abs/hep-lat/0605010} {arXiv:hep-lat/0605010} \BibitemShut {NoStop}%
\bibitem [{\citenamefont {Furnstahl}\ \emph {et~al.}(2015)\citenamefont {Furnstahl}, \citenamefont {Klco}, \citenamefont {Phillips},\ and\ \citenamefont {Wesolowski}}]{Furnstahl:2015rha}%
  \BibitemOpen
  \bibfield  {author} {\bibinfo {author} {\bibfnamefont {R.~J.}\ \bibnamefont {Furnstahl}}, \bibinfo {author} {\bibfnamefont {N.}~\bibnamefont {Klco}}, \bibinfo {author} {\bibfnamefont {D.~R.}\ \bibnamefont {Phillips}}, \ and\ \bibinfo {author} {\bibfnamefont {S.}~\bibnamefont {Wesolowski}},\ }\href {\doibase 10.1103/PhysRevC.92.024005} {\bibfield  {journal} {\bibinfo  {journal} {Phys. Rev. C}\ }\textbf {\bibinfo {volume} {92}},\ \bibinfo {pages} {024005} (\bibinfo {year} {2015})},\ \Eprint {http://arxiv.org/abs/1506.01343} {arXiv:1506.01343 [nucl-th]} \BibitemShut {NoStop}%
\bibitem [{\citenamefont {Melendez}\ \emph {et~al.}(2019)\citenamefont {Melendez}, \citenamefont {Furnstahl}, \citenamefont {Phillips}, \citenamefont {Pratola},\ and\ \citenamefont {Wesolowski}}]{Melendez:2019izc}%
  \BibitemOpen
  \bibfield  {author} {\bibinfo {author} {\bibfnamefont {J.~A.}\ \bibnamefont {Melendez}}, \bibinfo {author} {\bibfnamefont {R.~J.}\ \bibnamefont {Furnstahl}}, \bibinfo {author} {\bibfnamefont {D.~R.}\ \bibnamefont {Phillips}}, \bibinfo {author} {\bibfnamefont {M.~T.}\ \bibnamefont {Pratola}}, \ and\ \bibinfo {author} {\bibfnamefont {S.}~\bibnamefont {Wesolowski}},\ }\href {\doibase 10.1103/PhysRevC.100.044001} {\bibfield  {journal} {\bibinfo  {journal} {Phys. Rev. C}\ }\textbf {\bibinfo {volume} {100}},\ \bibinfo {pages} {044001} (\bibinfo {year} {2019})},\ \Eprint {http://arxiv.org/abs/1904.10581} {arXiv:1904.10581 [nucl-th]} \BibitemShut {NoStop}%
\bibitem [{\citenamefont {Melendez}\ \emph {et~al.}(2017)\citenamefont {Melendez}, \citenamefont {Wesolowski},\ and\ \citenamefont {Furnstahl}}]{Melendez:2017phj}%
  \BibitemOpen
  \bibfield  {author} {\bibinfo {author} {\bibfnamefont {J.~A.}\ \bibnamefont {Melendez}}, \bibinfo {author} {\bibfnamefont {S.}~\bibnamefont {Wesolowski}}, \ and\ \bibinfo {author} {\bibfnamefont {R.~J.}\ \bibnamefont {Furnstahl}},\ }\href {\doibase 10.1103/PhysRevC.96.024003} {\bibfield  {journal} {\bibinfo  {journal} {Phys. Rev. C}\ }\textbf {\bibinfo {volume} {96}},\ \bibinfo {pages} {024003} (\bibinfo {year} {2017})},\ \Eprint {http://arxiv.org/abs/1704.03308} {arXiv:1704.03308 [nucl-th]} \BibitemShut {NoStop}%
\bibitem [{\citenamefont {Wesolowski}\ \emph {et~al.}(2021)\citenamefont {Wesolowski}, \citenamefont {Svensson}, \citenamefont {Ekstr{\"o}m}, \citenamefont {Forss{\'e}n}, \citenamefont {Furnstahl}, \citenamefont {Melendez},\ and\ \citenamefont {Phillips}}]{Wesolowski:2021cni}%
  \BibitemOpen
  \bibfield  {author} {\bibinfo {author} {\bibfnamefont {S.}~\bibnamefont {Wesolowski}}, \bibinfo {author} {\bibfnamefont {I.}~\bibnamefont {Svensson}}, \bibinfo {author} {\bibfnamefont {A.}~\bibnamefont {Ekstr{\"o}m}}, \bibinfo {author} {\bibfnamefont {C.}~\bibnamefont {Forss{\'e}n}}, \bibinfo {author} {\bibfnamefont {R.~J.}\ \bibnamefont {Furnstahl}}, \bibinfo {author} {\bibfnamefont {J.~A.}\ \bibnamefont {Melendez}}, \ and\ \bibinfo {author} {\bibfnamefont {D.~R.}\ \bibnamefont {Phillips}},\ }\href {\doibase 10.1103/PhysRevC.104.064001} {\bibfield  {journal} {\bibinfo  {journal} {Phys. Rev. C}\ }\textbf {\bibinfo {volume} {104}},\ \bibinfo {pages} {064001} (\bibinfo {year} {2021})},\ \Eprint {http://arxiv.org/abs/2104.04441} {arXiv:2104.04441 [nucl-th]} \BibitemShut {NoStop}%
\bibitem [{\citenamefont {Drischler}\ \emph {et~al.}(2020)\citenamefont {Drischler}, \citenamefont {Furnstahl}, \citenamefont {Melendez},\ and\ \citenamefont {Phillips}}]{Drischler:2020hwi}%
  \BibitemOpen
  \bibfield  {author} {\bibinfo {author} {\bibfnamefont {C.}~\bibnamefont {Drischler}}, \bibinfo {author} {\bibfnamefont {R.~J.}\ \bibnamefont {Furnstahl}}, \bibinfo {author} {\bibfnamefont {J.~A.}\ \bibnamefont {Melendez}}, \ and\ \bibinfo {author} {\bibfnamefont {D.~R.}\ \bibnamefont {Phillips}},\ }\href {\doibase 10.1103/PhysRevLett.125.202702} {\bibfield  {journal} {\bibinfo  {journal} {Phys. Rev. Lett.}\ }\textbf {\bibinfo {volume} {125}},\ \bibinfo {pages} {202702} (\bibinfo {year} {2020})},\ \Eprint {http://arxiv.org/abs/2004.07232} {arXiv:2004.07232 [nucl-th]} \BibitemShut {NoStop}%
\bibitem [{\citenamefont {Millican}\ \emph {et~al.}(2024)\citenamefont {Millican}, \citenamefont {Furnstahl}, \citenamefont {Melendez}, \citenamefont {Phillips},\ and\ \citenamefont {Pratola}}]{Millican:2024yuz}%
  \BibitemOpen
  \bibfield  {author} {\bibinfo {author} {\bibfnamefont {P.~J.}\ \bibnamefont {Millican}}, \bibinfo {author} {\bibfnamefont {R.~J.}\ \bibnamefont {Furnstahl}}, \bibinfo {author} {\bibfnamefont {J.~A.}\ \bibnamefont {Melendez}}, \bibinfo {author} {\bibfnamefont {D.~R.}\ \bibnamefont {Phillips}}, \ and\ \bibinfo {author} {\bibfnamefont {M.~T.}\ \bibnamefont {Pratola}},\ }\href {\doibase 10.1103/PhysRevC.110.044002} {\bibfield  {journal} {\bibinfo  {journal} {Phys. Rev. C}\ }\textbf {\bibinfo {volume} {110}},\ \bibinfo {pages} {044002} (\bibinfo {year} {2024})},\ \Eprint {http://arxiv.org/abs/2402.13165} {arXiv:2402.13165 [nucl-th]} \BibitemShut {NoStop}%
\bibitem [{\citenamefont {Millican}\ \emph {et~al.}(2025)\citenamefont {Millican}, \citenamefont {Furnstahl}, \citenamefont {Melendez},\ and\ \citenamefont {Phillips}}]{Millican:2025sdp}%
  \BibitemOpen
  \bibfield  {author} {\bibinfo {author} {\bibfnamefont {P.~J.}\ \bibnamefont {Millican}}, \bibinfo {author} {\bibfnamefont {R.~J.}\ \bibnamefont {Furnstahl}}, \bibinfo {author} {\bibfnamefont {J.~A.}\ \bibnamefont {Melendez}}, \ and\ \bibinfo {author} {\bibfnamefont {D.~R.}\ \bibnamefont {Phillips}},\ }\href@noop {} {\  (\bibinfo {year} {2025})},\ \Eprint {http://arxiv.org/abs/2508.17558} {arXiv:2508.17558 [nucl-th]} \BibitemShut {NoStop}%
\bibitem [{\citenamefont {McClung}\ \emph {et~al.}(2025)\citenamefont {McClung}, \citenamefont {Elster},\ and\ \citenamefont {Phillips}}]{McClung:2025rtc}%
  \BibitemOpen
  \bibfield  {author} {\bibinfo {author} {\bibfnamefont {B.}~\bibnamefont {McClung}}, \bibinfo {author} {\bibfnamefont {C.}~\bibnamefont {Elster}}, \ and\ \bibinfo {author} {\bibfnamefont {D.~R.}\ \bibnamefont {Phillips}},\ }\href {\doibase 10.1103/jq4y-ydhk} {\bibfield  {journal} {\bibinfo  {journal} {Phys. Rev. C}\ }\textbf {\bibinfo {volume} {111}},\ \bibinfo {pages} {064002} (\bibinfo {year} {2025})},\ \Eprint {http://arxiv.org/abs/2501.09231} {arXiv:2501.09231 [nucl-th]} \BibitemShut {NoStop}%
\bibitem [{\citenamefont {Ekstr{\"o}m}\ and\ \citenamefont {Platter}(2025{\natexlab{a}})}]{Ekstrom:2024dqr}%
  \BibitemOpen
  \bibfield  {author} {\bibinfo {author} {\bibfnamefont {A.}~\bibnamefont {Ekstr{\"o}m}}\ and\ \bibinfo {author} {\bibfnamefont {L.}~\bibnamefont {Platter}},\ }\href {\doibase 10.1016/j.physletb.2024.139207} {\bibfield  {journal} {\bibinfo  {journal} {Phys. Lett. B}\ }\textbf {\bibinfo {volume} {860}},\ \bibinfo {pages} {139207} (\bibinfo {year} {2025}{\natexlab{a}})},\ \Eprint {http://arxiv.org/abs/2409.08197} {arXiv:2409.08197 [nucl-th]} \BibitemShut {NoStop}%
\bibitem [{\citenamefont {Ekstr{\"o}m}\ and\ \citenamefont {Platter}(2025{\natexlab{b}})}]{Ekstrom:2025ncs}%
  \BibitemOpen
  \bibfield  {author} {\bibinfo {author} {\bibfnamefont {A.}~\bibnamefont {Ekstr{\"o}m}}\ and\ \bibinfo {author} {\bibfnamefont {L.}~\bibnamefont {Platter}},\ }\href {\doibase 10.1103/4j2q-yvr9} {\bibfield  {journal} {\bibinfo  {journal} {Phys. Rev. C}\ }\textbf {\bibinfo {volume} {112}},\ \bibinfo {pages} {L031002} (\bibinfo {year} {2025}{\natexlab{b}})},\ \Eprint {http://arxiv.org/abs/2507.08700} {arXiv:2507.08700 [nucl-th]} \BibitemShut {NoStop}%
\bibitem [{\citenamefont {Bub}\ \emph {et~al.}(2025)\citenamefont {Bub}, \citenamefont {Piarulli}, \citenamefont {Furnstahl}, \citenamefont {Pastore},\ and\ \citenamefont {Phillips}}]{Bub:2024gyz}%
  \BibitemOpen
  \bibfield  {author} {\bibinfo {author} {\bibfnamefont {J.~M.}\ \bibnamefont {Bub}}, \bibinfo {author} {\bibfnamefont {M.}~\bibnamefont {Piarulli}}, \bibinfo {author} {\bibfnamefont {R.~J.}\ \bibnamefont {Furnstahl}}, \bibinfo {author} {\bibfnamefont {S.}~\bibnamefont {Pastore}}, \ and\ \bibinfo {author} {\bibfnamefont {D.~R.}\ \bibnamefont {Phillips}},\ }\href {\doibase 10.1103/PhysRevC.111.034005} {\bibfield  {journal} {\bibinfo  {journal} {Phys. Rev. C}\ }\textbf {\bibinfo {volume} {111}},\ \bibinfo {pages} {034005} (\bibinfo {year} {2025})},\ \Eprint {http://arxiv.org/abs/2408.02480} {arXiv:2408.02480 [nucl-th]} \BibitemShut {NoStop}%
\bibitem [{\citenamefont {Gasser}\ \emph {et~al.}(2002)\citenamefont {Gasser}, \citenamefont {Ivanov}, \citenamefont {Lipartia}, \citenamefont {Mojzis},\ and\ \citenamefont {Rusetsky}}]{Gasser_2002}%
  \BibitemOpen
  \bibfield  {author} {\bibinfo {author} {\bibfnamefont {J.}~\bibnamefont {Gasser}}, \bibinfo {author} {\bibfnamefont {M.}~\bibnamefont {Ivanov}}, \bibinfo {author} {\bibfnamefont {E.}~\bibnamefont {Lipartia}}, \bibinfo {author} {\bibfnamefont {M.}~\bibnamefont {Mojzis}}, \ and\ \bibinfo {author} {\bibfnamefont {A.}~\bibnamefont {Rusetsky}},\ }\href {\doibase 10.1007/s10052-002-1013-z} {\bibfield  {journal} {\bibinfo  {journal} {The European Physical Journal C}\ }\textbf {\bibinfo {volume} {26}},\ \bibinfo {pages} {13–34} (\bibinfo {year} {2002})}\BibitemShut {NoStop}%
\bibitem [{\citenamefont {Siemens}\ \emph {et~al.}(2016)\citenamefont {Siemens}, \citenamefont {Bernard}, \citenamefont {Epelbaum}, \citenamefont {Gasparyan}, \citenamefont {Krebs},\ and\ \citenamefont {Mei{\ss}ner}}]{Siemens:2016hdi}%
  \BibitemOpen
  \bibfield  {author} {\bibinfo {author} {\bibfnamefont {D.}~\bibnamefont {Siemens}}, \bibinfo {author} {\bibfnamefont {V.}~\bibnamefont {Bernard}}, \bibinfo {author} {\bibfnamefont {E.}~\bibnamefont {Epelbaum}}, \bibinfo {author} {\bibfnamefont {A.}~\bibnamefont {Gasparyan}}, \bibinfo {author} {\bibfnamefont {H.}~\bibnamefont {Krebs}}, \ and\ \bibinfo {author} {\bibfnamefont {U.-G.}\ \bibnamefont {Mei{\ss}ner}},\ }\href {\doibase 10.1103/PhysRevC.94.014620} {\bibfield  {journal} {\bibinfo  {journal} {Phys. Rev. C}\ }\textbf {\bibinfo {volume} {94}},\ \bibinfo {pages} {014620} (\bibinfo {year} {2016})},\ \Eprint {http://arxiv.org/abs/1602.02640} {arXiv:1602.02640 [nucl-th]} \BibitemShut {NoStop}%
\bibitem [{\citenamefont {Siemens}\ \emph {et~al.}(2017)\citenamefont {Siemens}, \citenamefont {Bernard}, \citenamefont {Epelbaum}, \citenamefont {Gasparyan}, \citenamefont {Krebs},\ and\ \citenamefont {Mei{\ss}ner}}]{Siemens:2017opr}%
  \BibitemOpen
  \bibfield  {author} {\bibinfo {author} {\bibfnamefont {D.}~\bibnamefont {Siemens}}, \bibinfo {author} {\bibfnamefont {V.}~\bibnamefont {Bernard}}, \bibinfo {author} {\bibfnamefont {E.}~\bibnamefont {Epelbaum}}, \bibinfo {author} {\bibfnamefont {A.~M.}\ \bibnamefont {Gasparyan}}, \bibinfo {author} {\bibfnamefont {H.}~\bibnamefont {Krebs}}, \ and\ \bibinfo {author} {\bibfnamefont {U.-G.}\ \bibnamefont {Mei{\ss}ner}},\ }\href {\doibase 10.1103/PhysRevC.96.055205} {\bibfield  {journal} {\bibinfo  {journal} {Phys. Rev. C}\ }\textbf {\bibinfo {volume} {96}},\ \bibinfo {pages} {055205} (\bibinfo {year} {2017})},\ \Eprint {http://arxiv.org/abs/1704.08988} {arXiv:1704.08988 [nucl-th]} \BibitemShut {NoStop}%
\bibitem [{\citenamefont {Bernard}\ \emph {et~al.}(2025)\citenamefont {Bernard}, \citenamefont {Gegelia}, \citenamefont {Ghosh},\ and\ \citenamefont {Mei\ss{}ner}}]{Bernard:2025gto}%
  \BibitemOpen
  \bibfield  {author} {\bibinfo {author} {\bibfnamefont {V.}~\bibnamefont {Bernard}}, \bibinfo {author} {\bibfnamefont {J.}~\bibnamefont {Gegelia}}, \bibinfo {author} {\bibfnamefont {S.}~\bibnamefont {Ghosh}}, \ and\ \bibinfo {author} {\bibfnamefont {U.-G.}\ \bibnamefont {Mei\ss{}ner}},\ }\href@noop {} {\  (\bibinfo {year} {2025})},\ \Eprint {http://arxiv.org/abs/2505.05941} {arXiv:2505.05941 [hep-ph]} \BibitemShut {NoStop}%
\bibitem [{\citenamefont {Hoferichter}\ \emph {et~al.}(2016)\citenamefont {Hoferichter}, \citenamefont {Ruiz~de Elvira}, \citenamefont {Kubis},\ and\ \citenamefont {Mei{\ss}ner}}]{Hoferichter:2015hva}%
  \BibitemOpen
  \bibfield  {author} {\bibinfo {author} {\bibfnamefont {M.}~\bibnamefont {Hoferichter}}, \bibinfo {author} {\bibfnamefont {J.}~\bibnamefont {Ruiz~de Elvira}}, \bibinfo {author} {\bibfnamefont {B.}~\bibnamefont {Kubis}}, \ and\ \bibinfo {author} {\bibfnamefont {U.-G.}\ \bibnamefont {Mei{\ss}ner}},\ }\href {\doibase 10.1016/j.physrep.2016.02.002} {\bibfield  {journal} {\bibinfo  {journal} {Phys. Rept.}\ }\textbf {\bibinfo {volume} {625}},\ \bibinfo {pages} {1} (\bibinfo {year} {2016})},\ \Eprint {http://arxiv.org/abs/1510.06039} {arXiv:1510.06039 [hep-ph]} \BibitemShut {NoStop}%
\bibitem [{\citenamefont {Mei{\ss}ner}\ and\ \citenamefont {Rusetsky}(2022)}]{Meissner:2022cbi}%
  \BibitemOpen
  \bibfield  {author} {\bibinfo {author} {\bibfnamefont {U.-G.}\ \bibnamefont {Mei{\ss}ner}}\ and\ \bibinfo {author} {\bibfnamefont {A.}~\bibnamefont {Rusetsky}},\ }\href {\doibase 10.1017/9781108689038} {\emph {\bibinfo {title} {{Effective Field Theories}}}}\ (\bibinfo  {publisher} {Cambridge University Press},\ \bibinfo {year} {2022})\BibitemShut {NoStop}%
\bibitem [{\citenamefont {Hoferichter}\ and\ \citenamefont {de~Elvira}(2025)}]{Hoferichter:2025ubp}%
  \BibitemOpen
  \bibfield  {author} {\bibinfo {author} {\bibfnamefont {M.}~\bibnamefont {Hoferichter}}\ and\ \bibinfo {author} {\bibfnamefont {J.~R.}\ \bibnamefont {de~Elvira}},\ }\enquote {\bibinfo {title} {{Nucleon mass: trace anomaly and $\sigma$-terms}},}\ \ (\bibinfo {year} {2025})\ \Eprint {http://arxiv.org/abs/2506.23902} {arXiv:2506.23902 [hep-ph]} \BibitemShut {NoStop}%
\bibitem [{\citenamefont {McGovern}\ and\ \citenamefont {Birse}(1999)}]{McGovern:1998tm}%
  \BibitemOpen
  \bibfield  {author} {\bibinfo {author} {\bibfnamefont {J.~A.}\ \bibnamefont {McGovern}}\ and\ \bibinfo {author} {\bibfnamefont {M.~C.}\ \bibnamefont {Birse}},\ }\href {\doibase 10.1016/S0370-2693(98)01550-0} {\bibfield  {journal} {\bibinfo  {journal} {Phys. Lett. B}\ }\textbf {\bibinfo {volume} {446}},\ \bibinfo {pages} {300} (\bibinfo {year} {1999})},\ \Eprint {http://arxiv.org/abs/hep-ph/9807384} {arXiv:hep-ph/9807384} \BibitemShut {NoStop}%
\bibitem [{\citenamefont {Liang}\ \emph {et~al.}(2025)\citenamefont {Liang}, \citenamefont {Chen}, \citenamefont {Guo}, \citenamefont {Guo},\ and\ \citenamefont {Yao}}]{Liang:2025cjd}%
  \BibitemOpen
  \bibfield  {author} {\bibinfo {author} {\bibfnamefont {Z.-R.}\ \bibnamefont {Liang}}, \bibinfo {author} {\bibfnamefont {H.-X.}\ \bibnamefont {Chen}}, \bibinfo {author} {\bibfnamefont {F.-K.}\ \bibnamefont {Guo}}, \bibinfo {author} {\bibfnamefont {Z.-H.}\ \bibnamefont {Guo}}, \ and\ \bibinfo {author} {\bibfnamefont {D.-L.}\ \bibnamefont {Yao}},\ }\href@noop {} {\  (\bibinfo {year} {2025})},\ \Eprint {http://arxiv.org/abs/2502.19168} {arXiv:2502.19168 [hep-ph]} \BibitemShut {NoStop}%
\bibitem [{\citenamefont {Schindler}\ \emph {et~al.}(2008)\citenamefont {Schindler}, \citenamefont {Djukanovic}, \citenamefont {Gegelia},\ and\ \citenamefont {Scherer}}]{Schindler:2007dr}%
  \BibitemOpen
  \bibfield  {author} {\bibinfo {author} {\bibfnamefont {M.~R.}\ \bibnamefont {Schindler}}, \bibinfo {author} {\bibfnamefont {D.}~\bibnamefont {Djukanovic}}, \bibinfo {author} {\bibfnamefont {J.}~\bibnamefont {Gegelia}}, \ and\ \bibinfo {author} {\bibfnamefont {S.}~\bibnamefont {Scherer}},\ }\href {\doibase 10.1016/j.nuclphysa.2008.01.023} {\bibfield  {journal} {\bibinfo  {journal} {Nucl. Phys. A}\ }\textbf {\bibinfo {volume} {803}},\ \bibinfo {pages} {68} (\bibinfo {year} {2008})},\ \bibinfo {note} {[Erratum: Nucl.Phys.A 1010, 122175 (2021)]},\ \Eprint {http://arxiv.org/abs/0707.4296} {arXiv:0707.4296 [hep-ph]} \BibitemShut {NoStop}%
\bibitem [{\citenamefont {Conrad}\ \emph {et~al.}(2024)\citenamefont {Conrad}, \citenamefont {Gasparyan},\ and\ \citenamefont {Epelbaum}}]{Conrad:2024phd}%
  \BibitemOpen
  \bibfield  {author} {\bibinfo {author} {\bibfnamefont {N.~D.}\ \bibnamefont {Conrad}}, \bibinfo {author} {\bibfnamefont {A.}~\bibnamefont {Gasparyan}}, \ and\ \bibinfo {author} {\bibfnamefont {E.}~\bibnamefont {Epelbaum}},\ }\href {\doibase 10.22323/1.413.0074} {\bibfield  {journal} {\bibinfo  {journal} {PoS}\ }\textbf {\bibinfo {volume} {CD2021}},\ \bibinfo {pages} {074} (\bibinfo {year} {2024})}\BibitemShut {NoStop}%
\bibitem [{\citenamefont {Kambor}\ and\ \citenamefont {Mojzis}(1999)}]{Kambor:1998pi}%
  \BibitemOpen
  \bibfield  {author} {\bibinfo {author} {\bibfnamefont {J.}~\bibnamefont {Kambor}}\ and\ \bibinfo {author} {\bibfnamefont {M.}~\bibnamefont {Mojzis}},\ }\href {\doibase 10.1088/1126-6708/1999/04/031} {\bibfield  {journal} {\bibinfo  {journal} {JHEP}\ }\textbf {\bibinfo {volume} {04}},\ \bibinfo {pages} {031} (\bibinfo {year} {1999})},\ \Eprint {http://arxiv.org/abs/hep-ph/9901235} {arXiv:hep-ph/9901235} \BibitemShut {NoStop}%
\bibitem [{\citenamefont {Thim}\ \emph {et~al.}(2024)\citenamefont {Thim}, \citenamefont {Ekstr{\"o}m},\ and\ \citenamefont {Forss{\'e}n}}]{Thim:2024yks}%
  \BibitemOpen
  \bibfield  {author} {\bibinfo {author} {\bibfnamefont {O.}~\bibnamefont {Thim}}, \bibinfo {author} {\bibfnamefont {A.}~\bibnamefont {Ekstr{\"o}m}}, \ and\ \bibinfo {author} {\bibfnamefont {C.}~\bibnamefont {Forss{\'e}n}},\ }\href {\doibase 10.1103/PhysRevC.109.064001} {\bibfield  {journal} {\bibinfo  {journal} {Phys. Rev. C}\ }\textbf {\bibinfo {volume} {109}},\ \bibinfo {pages} {064001} (\bibinfo {year} {2024})},\ \Eprint {http://arxiv.org/abs/2402.15325} {arXiv:2402.15325 [nucl-th]} \BibitemShut {NoStop}%
\end{thebibliography}%

\end{document}